\documentstyle[eqsecnum,aps,preprint]{revtex}
\begin{document}
\draft
\title{The Real Symplectic Groups in \\ Quantum Mechanics and Optics}
\author{Arvind\cite{email}}
\address{Department of Physics\\
Indian Institute of Science, Bangalore - 560 012, India}
\author{B. Dutta}
\address{Jawaharlal Nehru Centre for Advanced Scientific Research\\
Jakkur, Bangalore - 560 064, India}
\author{N. Mukunda\cite{jncasr}}
\address{Centre for Theoretical Studies and Department of Physics\\
Indian Institute of Science, Bangalore - 560 012, India}
\author{and \\ R. Simon}
\address{Institute of Mathematical Sciences,
C. I. T. Campus, Madras - 600 113}
\date{\today}
\maketitle
\newpage
\begin{abstract}
We present a utilitarian review of the family of matrix groups
$Sp(2n,\Re)\/$, in a form suited to various applications both in
optics and quantum mechanics. We contrast these groups and their
geometry with the much more familiar Euclidean and unitary
geometries. Both the properties of finite group elements and of the
Lie algebra are studied, and special attention is paid to the
so-called unitary metaplectic representation of $Sp(2n,\Re)\/$.
Global decomposition theorems, interesting subgroups and their
generators are described. Turning to $n$-mode quantum systems, we
define and study their variance matrices in general states, the
implications of the Heisenberg uncertainty principles, and develop a
$U(n)$-invariant squeezing criterion. The particular properties of
Wigner distributions and Gaussian pure state wavefunctions under
$Sp(2n,\Re)\/$ action are delineated.
\end{abstract}
\section{Introduction}
The symplectic groups form one of the three major families of
classical semisimple Lie groups, the other two being the real
orthogonal family and the complex unitary family~\cite{families}.
Apart from the groups describing nonrelativistic and relativistic
space-time geometries, namely the Galilei, Lorentz and Poincare
groups, most of the Lie groups encountered in physical problems, for
example as symmetry groups, belong to either the real orthogonal or
the unitary families~\cite{ortho-uni}.  These are multidimensional as
well as complex generalisations of the rotation group of Euclidean
geometry characterising physical three-dimensional space.  As a
result, the intuitive geometrical ideas that go with real orthogonal
or complex unitary geometries are quite familiar to most physicists.

It has been realised recently, however, that in several problems both
in quantum mechanics and in optics, the real symplectic groups play
an important role~\cite{role}.  In the latter context, this is so in
both classical and quantum theories.  More generally, these groups
come in very naturally through the canonical formalism of classical
dynamics, and its counterpart in quantum mechanics.

Symplectic geometry, on the other hand, differs profoundly from the
real or complex Euclidean variety~\cite{geometry}.  Here the
intuitively familiar concepts of length, angle, perpendicularity and
the Pythagoras Theorem are all absent.  In their place we have
typically new concepts characteristic of canonical mechanics.

The purpose of this informal and utilitarian survey is to introduce
methods based on the real symplectic groups to those who are
otherwise familiar with the structures of quantum mechanics and/or
the theory of partial coherence in optics.  The intention is to
``inform rather than astonish'', and to describe the main features of
the symplectic point of view.  Our account will not contain complete
proofs of all statements presented, but a motivated reader should be
easily able to supply additional detail and proceed to make practical
use of these methods.

The material of this article is organised as follows.  Section 2
introduces the group $Sp(2n,\Re)\/$ as the group of linear
transformations preserving the classical Poisson Brackets as well as
the quantum commutation relations among $n$ pairs of canonical
variables. In the quantum case, the Stone - von Neumann theorem
allows us to infer that these transformations are unitarily
implementable. Section 3 develops some ideas related to real
symplectic linear vector spaces, specially the concepts of symplectic
complement and symplectic rank of a subspace, in order to contrast
symplectic geometry with real Euclidean and complex unitary
geometries. Some useful properties of the matrices occurring in the
defining representation of $Sp(2n,\Re)\/$, and their complex form,
are then explained in Section 4. Here we also list several useful
subgroups of $Sp(2n,\Re)\/$, and describe four global decomposition
theorems - the polar, Euler, pre -Iwasawa and Iwasawa decompositions.
Section 5 studies the Lie algebra $\underline{Sp(2n,\Re)}\/$, first
in the defining representation and then in a general, possibly
unitary, representation.  Convenient ways of breaking up the
generators into subsets, and the generators of various subgroups, are
described. In Section 6 we set up and study the special unitary
metaplectic representation of $Sp(2n,\Re)\/$ and relate it to the
generalised Huyghens kernel in any number of dimensions. The
characteristic differences between the compact and the noncompact
generators of $Sp(2n,\Re)\/$ are seen in their dependences on mode
annihilation and creation operators. The former(latter) conserve(do
not conserve) the total number operator. Section 7 studies the
relationship between the metaplectic unitary representation of
$Sp(2n,\Re)\/$, and two often used descriptions of quantum mechanical
operators, namely the Wigner function representation and the diagonal
coherent state representation. While the former is covariant, i.e,
transforms simply, under the full group $Sp(2n,\Re)\/$, the latter is
covariant only under the maximal compact subgroup $K(n)=U(n)$ of
$Sp(2n,\Re)\/$. Section 8 takes up the questions of defining the
noise or variance matrix for any state of an $n$-mode quantum system,
both in real and complex forms, and their behaviours under
$Sp(2n,\Re)\/$. In Section 9 we carry this analysis further to show
that the Heisenberg uncertainty principles for any number of modes
can be given in explicitly $Sp(2n,\Re)\/$ covariant forms; a key role
here is played by Williamson's Theorem relating to normal forms of
quadratic Hamiltonians. This study leads in Section 10 to the setting
up of an $U(n)$-invariant squeezing criterion for $n$-mode systems.
This is the maximal physically reasonable invariance one could ask
for in these systems, and it can be stated very elegantly in terms of
the general variance matrix set up in Section 8. Section 11
describes, and motivates, some interesting classes of variance
matrices with distinctive group theoretic properties, and Section 12
is devoted to a study of general centred and normalized pure Gaussian
wavefunctions for $n$-mode systems. The transitive action of
$Sp(2n,\Re)\/$ on these wavefunctions, via the symplectic
representation, and the emergence of a matrix form of the Mobius
transformation, are described. Section 13 contains some concluding
remarks.

\section{The Real Symplectic Groups $S\lowercase{p}(2\lowercase{n},\Re)$}

We consider a classical or quantum canonical system with n degrees of
freedom, that is, n pairs of mutually conjugate canonical variables.
In the classical case these are numerical variables written as $q_r,
p_r, r= 1,2, \ldots, n$.  In quantum mechanics we have an irreducible
set of hermitian operators $\hat{q}_{r}, \hat{p}_{r}$ acting on a
suitable Hilbert space ${\cal H}$.  The basic kinematic structure is
given by Poisson brackets (PB) in one case, by the Heisenberg
commutation relations (CR) in the other.  To express them both
compactly and elegantly, we introduce the following notation.  We
assemble the $q$'s and $p$'s into $2n$-component column vectors $\xi,
\hat{\xi}$:
\begin{eqnarray}
\xi  & = &  (\xi_{a})  = (q_{1} \ldots q_n\/ p_{1} \ldots p_{n})^{T},
\nonumber \\
\hat{\xi} & = & (\hat{\xi}_{a}) = (\hat{q}_{1} \ldots
\hat{q}_{n}\/ \hat{p}_{1} \ldots\hat{p}_{n})^{T}, \,a=1,2, \ldots, 2n
\end{eqnarray}
Then the classical PB's and the quantum CR's are, respectively:
\begin{eqnarray}
\{ \xi_{a}, \xi_{b} \} & = & \beta_{ab},
\nonumber \\
\mbox{}[\hat{\xi}_{a}, \hat{\xi}_{b}] & = & i\hbar \beta_{ab},
\nonumber \\
\beta  & = & (\beta_{ab}) =
\left( \begin{array}{ll}
0_{n\times n} & 1_{n\times n}\\ -1_{n\times n} & 0_{n\times n}
\end{array} \right)
\end{eqnarray}
The usual relations stated separately in terms of $q$'s and $p$'s are
all contained here; and the even-dimensional real antisymmetric
matrix $\beta$ will play an important role.

We assume always that all the $\xi_{a}, \hat{\xi}_{a}$ are of
Cartesian type: the natural range (spectrum) for each of them is the
entire real line $\Re$

Changes of $\xi, \hat{\xi}$ to new quantities $\xi^\prime,
\hat{\xi}^\prime$ given as functions of the old ones such that
the basic kinematic relations are preserved may be called canonical
transformations in both situations:
\begin{eqnarray}
\xi^\prime &=& \mbox{numerical functions of}\, \xi  :
\{\xi^\prime_{a}, \xi^\prime_{b} \} =\beta_{ab}; \nonumber \\
\hat{\xi}^\prime& = & \mbox{operator functions of}\, \hat{\xi}:
[\hat{\xi}^\prime_{a}, \hat{\xi}^\prime_{b}] = i\hbar \beta_{ab}
\end{eqnarray}
Apart from C-number translations (shift of origin) the simplest such
transformations are the linear homogeneous ones.  Each such
transformation may be specified by a real 2n-dimensional matrix $S$,
the actions being
\begin{eqnarray}
S= (S_{ab}):\quad \xi^\prime_{a} & = & S_{ab}\xi_{b}, \nonumber \\
\hat{\xi}^\prime_{a} & = & S_{ab}  \hat{\xi}_{b}
\end{eqnarray}
In either case, the requirements (2.3) lead to a matrix condition on
$S$~~\cite{families}:
\begin{equation}
S \beta S^{T} = \beta
\end{equation}
This is the defining condition for the real symplectic group in $2n$
dimensions:
\begin{equation}
Sp(2n, \Re) = \{\mbox{$S =\;$real $2n\times 2n$ matrix}\,
\mid S\beta S^{T} =\beta \}
\end{equation}
The matrix $\beta$ is real, even-dimensional, antisymmetric and
nonsingular.  It is a ``symplectic metric matrix''.  As we see
explicitly later, $Sp(2n, \Re)$ transformations preserve symplectic
scalar products and the symplectic metric.

In quantum mechanics the Hilbert space ${\cal H}$ on which the
$\hat{\xi}_{a}$ act irreducibly can be described in many ways.  The
most familiar is the Schr\"{o}dinger description using wave functions
on ${\Re}^{n}$, that is, elements of L$^{2}(\Re^{n})$.  The
$\hat{q}_{r}$ act multiplicatively while the $\hat{p}_{r}$ are
differential operators:
\begin{eqnarray}
{\cal H} & = & \{ \psi(\b{$q$}) \mid \int_{{\Re}^{n}} d^{n}q \mid
\psi(\b{$q$})\mid^{2}  <  \infty \};
\nonumber \\
(\hat{q}_{r}\psi) (\b{$q$}) & = & q_{r} \psi(\b{$q$}),
\nonumber \\
(\hat{p}_{r}\psi) (\b{$q$}) & = & -i \hbar\frac{\partial}{\partial
q_{r}} \psi(\b{$q$}),
\nonumber \\
\b{$q$}  & = &(q_{1}, \ldots, q_{n})  \in {\Re}^{n}
\end{eqnarray}
Since the $\hat{\xi}_{a}$ are hermitian and irreducible, and since
for any $S \in Sp (2n, \Re)$ the transformed $\hat{\xi}^\prime_{a}$
are also hermitian and irreducible and obey the same CR's, by the
Stone-von Neumann theorem~\cite{stone-von} the change $\hat{\xi}
\rightarrow \hat{\xi}^\prime$ is  unitarily implementable.
Thus for each $S \in Sp(2n, \Re)$ it is definitely possible to
construct a unitary operator ${\cal U}(S)$ acting on ${\cal H}$ such
that
\begin{eqnarray}
\hat{\xi}^\prime_{a}  =  S_{ab}  \hat{\xi}_{b} & = &
{\cal U}(S)^{-1} \hat{\xi}_{a}\/{\cal U}(S),
\nonumber \\
{\cal U}(S)^{\dag} {\cal U}(S) &=& 1 \quad \mbox{on}\quad {\cal H}
\end{eqnarray}
This ${\cal U}(S)$ is arbitrary upto an $S$-dependent phase factor.
The general composition law that follows from the irreducibility of
the $\hat{\xi}_{a}$ is:
\begin{equation}
S^\prime, S \in Sp(2n, \Re)\,:\,{\cal U}(S^\prime) {\cal U}(S) =
(\mbox{phase factor dependent on}\,\, S^\prime, S) {\cal U}(S^\prime
S)
\end{equation}
We shall discuss in Section 6 the maximum simplification that can be
achieved in this phase factor by exploiting the phase freedom in each
${\cal U}(S)$.

\section{Aspects of symplectic geometry}

In this Section we develop a few basic concepts related to symplectic
vector spaces, so that the contrast with Euclidean and unitary
geometries can be clearly seen~\cite{geometry}.

Let $V$ be a real $2n$-dimensional vector space, with vectors ${x,y,
\cdots}$.  Suppose a nondegenerate bilinear
antisymmetric form (.,.) - a ``scalar product'' - is given on $V$.
Thus for any vectors $x, y \in V, \; (x,y)$ is a real number
separately linear in $x$ and $y$; and in addition the following hold:
\begin{eqnarray}\mbox{antisymmetry:}\quad (x,y) & = & -(y,x),
\nonumber \\
\mbox{nondegeneracy:}\quad(x,y) & = & 0 \quad \mbox{for all}
\; y \quad \Longleftrightarrow \quad x = 0
\end{eqnarray}
Then $V$ is a symplectic vector space, and (.,.) is a symplectic
scalar product.

We have stated the properties of the bilinear form in a basis
independent way.  It can be shown that if in a general basis we
express $(x,y)$ in terms of components of $x$ and $y$ in the form
\begin{equation}
(x,y) = x^{T} \eta y,
\end{equation}
involving an antisymmetric nonsingular matrix $\eta$, we can always
change to more convenient bases in which $\eta$ takes on particularly
simple canonical, or normal, forms.  Two such forms are worth
mentioning.  In one, $\eta$ becomes the matrix $\beta$ of eq.(2.2):
\begin{equation}
\eta = \beta: (x,y) = x_{1} y_{n+1} + x_{2}
y_{n+2} + \cdots + x_{n} y_{2n} - x_{n+1} y_{1} - x_{n+2} y_{2} -
\cdots  -  x_{2n} y_{n}
\end{equation}
Here the first and $(n+1){th}$ components belong to one canonical
pair; the second and $(n+2){th}$ to the second pair; and so on.
Another normal form disposes of the canonical pairs one at a time :
\begin{eqnarray}
\eta  =  \mbox{block-diag}(i \sigma_{2}, i \sigma_{2},
\cdots, i \sigma_{2}):
\nonumber \\
(x,y) = x_{1}y_{2} - x_{2}y_{1}+ x_{3}y_{4}-x_{4}y_{3} +
\cdots
\end{eqnarray}
With the normal form (3.3) the meaning of the defining condition
(2.5) for symplectic matrices becomes geometrically clear:
\begin{equation}
S \in Sp(2n, \Re), x^\prime = Sx\quad ,\quad y^\prime = Sy
\Rightarrow (x^\prime,  y^\prime)  = (x,y)
\end{equation}
In this sense the symplectic scalar product is preserved.  Of course
this means that we could have replaced the condition (2.5) by another
entirely equivalent one, using in place of $ \beta $ the matrix in
eq.(3.4).

Now let us look at linear subspaces $V_{1} \subseteq V$.  In both
Euclidean and unitary geometries it is well known that all subspaces
of the same dimension are basically similar, and cannot be
distinguished from each other in any intrinsic sense.  With
symplectic geometry there is a difference, as a new concept comes in.

Given a subspace ${V}_{1}$, we consider the bilinear form $(x,y)$
defined over $V$, restrict both arguments to $V_{1}$, and regard the
result as a bilinear form on $V_{1}$.  Now the nondegeneracy property
may well fail! Thus, there may exist a vector $x\in V_1\/$ such that
$(x,y)=0\/$ for all $y\in V_1$.  So we define the rank of this
restricted form as the symplectic rank of $V_{1}$:
\begin{equation}
\mbox{Symp. rk.} \;V_{1}  = \mbox{rank}\; (x,y),
\quad x\; \mbox{and}\; y  \in V_{1}
\end{equation}
Thus, if $x_r,\; r=1,\cdots k\/$ is a basis for $V_1\/$, where $k\/$
is the dimension of $V_1\/$, then Symp. rk. $V_1\/$ is the rank of
the $k\times k\/$ antisymmetric matrix $\left( (x_r,x_s) \right)$.
The symplectic rank is necessarily an even integer.  We have the
obvious limits
\begin{equation}
0 \;\leq \mbox{symp. rk.} \; V_{1} \; \leq k
\end{equation}
But nondegeneracy of (.,.) over $V$ leads to another nontrivial lower
bound, which is effective if $k > n$:
\begin{equation}
(0, 2(k-n))_{>} \leq \; \mbox{symp. rk.} \; V_{1}\; \leq k
\end{equation}
Basically we can say that the symplectic rank of a subspace $V_{1}$
is twice the number of complete canonical pairs contained in $V_{1}$.
Clearly this concept is symplectic invariant.  In particular two
subspaces $V_{1}$ and $V_{1}^\prime$ of the same dimension cannot be
mapped on to one another by any $Sp(2n, \Re)$ element if they have
unequal symplectic ranks.

As in the Euclidean case, we can pass from $V_{1}$ to its complement
written for convenience as $V_{1}^{\perp}$.  But the geometrical
significance is quite different. We call $V_{1}^{\perp}$ the
symplectic complement of $V_{1}$ and define it as a subspace of $V$
by:
\begin{equation}
V_{1}^{\perp} = \{ x \in V \mid (x,y) = 0 \quad\mbox{for all} \quad y
\in  V_{1} \}
\end{equation}
Taking the complement twice gives back $V_{1}$:
\begin{equation}(V_{1}^{\perp})^{\perp}  =  V_{1}
\end{equation}
This is as in the Euclidean case.  Even the dimensions follow the
same rule:
\begin{equation}
{\rm Dim} V_{1}^{\perp} = 2n - k
\end{equation}
This can be shown by using the nondegeneracy of (.,.).  But there the
similarity ends.  It can well happen that $V_{1}$ and $V_{1}^{\perp}$
have nontrivial intersection, and to that extent their sum does not
give back all of $V$. In general,
\begin{eqnarray}
V_{1} &\cap& V_{1}^{\perp} \neq 0,
\nonumber \\
V &\neq& V_{1} \oplus V_{1}^{\perp}
\end{eqnarray}
The two symplectic ranks can be related:
\begin{equation}
\mbox{Symp. rk.} \; V_{1}^{\perp} = 2(n-k) + \mbox{symp.
rk.}\; V_{1}
\end{equation}

The extreme case of eq.(3.12), which is very nonintuitive on the
basis of Euclidean geometric notions, is when symp. rk. $V_{1}$
vanishes.  For this case we give a special name and find:
\begin{eqnarray}
\mbox{Symp. rk.}\; V_{1} = 0  \Longleftrightarrow \nonumber \\
V_{1} \quad \mbox{is an isotropic subspace of} \quad V
\Longleftrightarrow
\nonumber \\
(x,y) = 0 \quad\mbox{for all}\quad x,y \in V_{1} \Longleftrightarrow
\nonumber \\
V_{1} \subseteq V_{1}^{\perp} \Rightarrow k \leq n
\end{eqnarray}
So in this case $V_{1}$ is contained in $V_{1}^{\perp}$.  The
opposite can also happen and then we call $V_{1}$ a co-isotropic
subspace:
\begin{eqnarray}
V_{1} \quad \mbox{is a co-isotropic subspace of}\; V
\Longleftrightarrow
\nonumber \\
V_{1}^{\perp} \quad \mbox{is isotropic} \Longleftrightarrow
\nonumber \\
V_{1} \supseteq V_{1}^{\perp} \Longrightarrow k \geq n
\end{eqnarray}
So if one of the pair $V_{1}, V_{1}^{\perp}$ is isotropic, the other
is co-isotropic.

An isotropic subspace has dimension $k \leq n$, while a co-isotropic
one has dimension $k \geq n$.  When they coincide, we have a special
situation and name.  An n-dimensional subspace $V_{1}$ of $V\/$ which
has vanishing symplectic rank is both isotropic and co-isotropic, and
coincides with $V_{1}^{\perp}$.  It is called a Lagrangian subspace.
This notion is important in Hamilton-Jacobi theory in classical
dynamics; it is also relevant in the choice of complete commuting
sets of operators in quantum mechanics.

These properties and notions give a feeling for symplectic geometry,
and for the ways in which it differs from orthogonal and unitary
geometries.  In particular the notions of length, angle and
perpendicularity are no longer available.

\section{Properties of $S\lowercase{p}(2\lowercase{n},\Re)$
matrices, complex form, subgroups, decompositions}

The matrices in the defining representation of $Sp(2n,\Re)$ obey
eq.(2.5).  From here many useful consequences follow, and we list
them:
\newcounter{a}
\begin{list}{(\roman{a})}{\usecounter{a}}
\item  $Sp(2n,\Re)$ is of dimension $n(2n+1)$.
\item $\beta  \in  Sp(2n,\Re)$.
\item $S  \in  Sp(2n,\Re)  \Rightarrow  -S, S^{-1}, S^{T}
\in  Sp(2n,\Re)$,\\
$S^{T} = \beta S^{-1} \beta^{-1}, (S^{-1})^{T} = \beta S \beta^{-1},
S^{-1} = \beta S^{T} \beta^{-1}$.
\item $det S  = +1$.
\item $S  \in Sp(2n,\Re)  \Rightarrow$ eigenvalue spectrum of
$S$ is invariant under reflection about the real axis, and through
unit circle ($ r e^{i\theta} \rightarrow \frac{1}{r} e^{i \theta}$);
eigenvalues $\pm 1$ have even multiplicities.
\begin{equation}
\end{equation}
\end{list}
Property (i) can be seen from the number of conditions contained in
eq.(2.5), and will be confirmed at the Lie algebra level.  While
properties (ii) and (iii) are easily checked, (iv) is rather subtle;
an indication will be given later to obtain it.  Property (v) is a
consequence of $S$ and $(S^{-1})^{T}$ being real and related by a
similarity transformation.

Sometimes it is convenient to write $S$ in $n \times n$ block form,
and then eq.(2.5) becomes a set of conditions on the blocks:
\begin{eqnarray}
S & = & \left( \begin{array}{ll} A & B \\ C & D \end{array} \right)
\in  Sp(2n,\Re):
\nonumber \\
S \beta S^{T} & = & \beta \Longleftrightarrow AB^{T}, CD^{T} \quad
{\rm symmetric}, AD^{T} - BC^{T} = 1_{n\times n}
\nonumber \\
S^{T} \beta S & = & \beta \Longleftrightarrow A^{T}C, B^{T}D \quad
{\rm symmetric}\,, A^{T}D - C^{T}B = 1_{n \times n}
\end{eqnarray}
While it is easy to check (as mentioned above) that $S \in
Sp(2n,\Re)$ implies $S^{T} \in Sp(2n,\Re)$ as well, it is not so easy
to pass directly from the first set of conditions above to the second
set, aside from reconstituting $A, B, C, D$ into $S$ and then passing
to $S^{T}$!

\underline{Complex form of $Sp(2n,\Re)$}

The $\beta$ matrix reflects the precise way in which the real $q$'s
and $p$'s have been put together in eq. (2.1) to form the $2n$
component object $\xi$ with real entries.  Sometimes it is
convenient, for instance in dealing with modes of the radiation
field, to work with complex combinations of the $\hat{q}$'s and
$\hat{p}$'s - mode annihilation and creation operators defined in
this way:
\begin{equation}
\hat{a}_{j}  =  \frac{1}{\sqrt 2}  (\hat{q}_{j} + i
\hat{p}_{j}), \hat{a}_{j}^{\dag}  = \frac{1}{\sqrt 2}
(\hat{q}_{j} - i \hat{p}_{j}), j=1, \cdots , n
\end{equation}
It is useful to arrange these into a new column vector
$\hat{\xi}^{(c)}$ with non-hermitian entries,
\begin{eqnarray}
\hat{\xi}^{(c)}  &=&  (\hat{\xi}^{(c)}_{a}) = (\hat{a}_{1}
\cdots \hat{a}_{n}\quad \hat{a}_{1}^{\dag} \cdots
\hat{a}_{n}^{\dag})^{T}  =  \Omega \hat{\xi},
\nonumber \\
\hat{\xi}  &=& \Omega^{\dag} \hat{\xi}^{(c)},
\nonumber \\
\Omega  &=&  \frac{1}{\sqrt{2}} \left(
\begin{array}{cc} 1 & i1 \\ 1 &-i1 \end{array} \right),
\Omega^{-1}  =  \Omega^{\dag} =  \frac{1}{\sqrt{2}}
\left( \begin{array}{cc} 1 & 1 \\ -i1 & i1 \end{array} \right)
\end{eqnarray}
Then the basic commutation relations in (2.2) can be written in two
equivalent ways:
\begin{eqnarray}
[ \hat{\xi}^{(c)}_{a}, \hat{\xi}^{(c)}_{b} ] &=& \beta_{ab},
\nonumber \\
\mbox{}[ \hat{\xi}^{(c)}_{a}, \hat{\xi}^{(c)\dag}_{b} ] &=&
(\Sigma_{3})_{ab},
\nonumber \\
\Sigma_{3}  &=&  \left( \begin{array}{cc} 1 & 0 \\ 0 & -1
\end{array} \right)
\end{eqnarray}
Now when we subject $\hat{\xi}$ to the real transformation $S \in
Sp(2n,\Re), \hat{\xi}^{(c)}$ experiences an equivalent complex
transformation :
\begin{eqnarray}
\hat{\xi}^\prime & = & S \; \hat{\xi} \quad  \Longleftrightarrow \quad
\hat{\xi}^{\prime (c)}  =  S^{(c)} \;  \hat{\xi}^{(c)}, \nonumber \\
S^{(c)} & = & \Omega \; S \; \Omega^{-1} \nonumber \\ & = &
\frac{1}{2} \left( \begin{array}{cc} A+D+i(C-B) &
\quad A-D+i(B+C)\\
A-D-i(B+C) & \quad A+D+i(B-C) \end{array} \right).  \end{eqnarray}
Thus $S^{(c)}$ is just a convenient complex form of the real
transformation S, much like the passage from Cartesian to spherical
components of spherical tensors.

\underline{Some Subgroups of $Sp(2n,\Re)$}

We shall describe here some useful subgroups of $Sp(2n,\Re)$.  Their
dimensions will be given, and where it is useful their complex forms
exhibited

\newcounter{listnumber}
\begin{list}{(\alph{listnumber})}{\usecounter{listnumber}}
\item $GL(n,\Re)\/$: This is the n$^{2}$ -dimensional general real
linear group; in terms of the block matrices $A, B, C, D\/$ it is
given thus :
\begin{equation}
A \in GL(n,\Re), B =C = 0, D = (A^{-1})^{T}
\end{equation}
Here the $\hat{q}$'s are subject to a general real linear nonsingular
transformation among themselves, and then the $\hat{p}$'s change in a
compensating contragredient manner.
\item $O(n,\Re)$: This is the orthogonal subgroup of $GL(n,\Re)$,
of dimension $\frac{1}{2}n(n-1)$.  It is that part of $GL(n,\Re)$
under which the $\hat{q}$'s and the $\hat{p}$'s change in the same
way:
\begin{equation}
A = D \in O(n,\Re), B = C = 0
\end{equation}
If we impose the condition $det A = +1$, we get the subgroup
$SO(n,\Re)\/$ of proper orthogonal transformations.
\item $U(n)\/$: Now we come to the $n$-dimensional unitary group, of
dimension $n^2$, a maximal compact subgroup within the noncompact
$Sp(2n,\Re)$.  We shall sometimes write K(n), or simply K, for it.
The corresponding symplectic matrices $S$ are identified as follows.
If we split any $U \in U(n)$ into real and imaginary parts we find
the properties
\begin{eqnarray}
U = X-iY \in U(n)\;,\quad U^{\dag}U =UU^{\dag} = 1
\Longleftrightarrow
\nonumber \\
X^{T}X + Y^{T}Y = XX^{T} + YY^{T} = 1,
\nonumber \\
X^{T}Y\;,\; XY^{T}\quad {\rm symmetric}
\end{eqnarray}
We can then produce a solution to the matrix condition (4.2)!  We
find:
\begin{eqnarray}
A = D & = & X, B = -C = Y
\nonumber \\
S(X,Y) & = & \left(\begin{array}{cc} X & Y\\-Y & X \end{array}\right)
\in  Sp(2n,\Re)
\end{eqnarray}
It is an interesting and easy exercise to check the following: If a
$2n \times 2n$ real matrix is both orthogonal and symplectic, then it
is unimodular as well and has to have the form $S(X,Y)$ for some
$U=X-iY \in U(n):$
\begin{eqnarray}
O(2n, \Re) &\cap& Sp(2n, \Re) = SO(2n, \Re) \cap Sp(2n,\Re)
\nonumber \\
=K(n) &=& \{ S(X,Y) \mid X-iY \in U(n) \}
\end{eqnarray}
The complex form of these matrices is very revealing:
\begin{eqnarray}
S^{(c)}(X,Y) & = & \Omega S(X,Y) \Omega^{-1}
\nonumber \\
&=& S^{(c)}(U)
\nonumber \\
&=& \left( \begin{array}{cc} U & 0 \\ 0 & U^{*} \end{array} \right)
\end{eqnarray}
So the $\hat{a}$'s and the $\hat{a}^{\dag}$'s undergo separate
unitary rotations, not mixing with one another:
\begin{equation}
U \in U(n):\quad \hat{a} \rightarrow U\hat{a},\quad
\hat{a}^{\dag}  \rightarrow  U^{*}\hat{a}^{\dag}
\end{equation}
Indeed , $K(n)$ is the maximal subgroup of $Sp(2n,\Re)$ such that
$\hat{a}$'s and $\hat{a}^{\dag}$'s transform independently.  We also
have the expected relation between the subgroups $O(n,\Re),
GL(n,\Re), {\rm and}\; U(n)$ exhibited above:
\begin{equation}
O(n,\Re) = GL(n,\Re) \cap U(n)
\end{equation}
Finally we turn to some Abelian subgroups~\cite{subgroups}.
\item $T^{f}$: This is a subgroup of dimension $\frac{1}{2}
n(n+1)$ and may be called the ``free propagation'' subgroup:
\begin{equation}A  = D  = 1, B =B^{T},  C  = 0
\end{equation}
The name comes from the actions on $\hat{q}$ and on $\hat{p}$:
\begin{equation}
\hat{q}^\prime  = \hat{q}  + B\hat{p}, \quad  \hat{p}^\prime  =
\hat{p}
\end{equation}
Group composition corresponds to adding the B matrices, which
explains the Abelian nature.
\item $T^{(l)}$:  This is the result of conjugating elements
of $T^{(f)}$ by $\beta$.  We call it the ``lens'' subgroup, on
account of the action on $\hat{q}$'s and $\hat{p}$'s:
\[
A = D = 1, B =0, C = C^{T};
\]
\begin{equation}
\hat{q}^\prime  = \hat{q}, \quad  \hat{p}^\prime
= \hat{p} + C\hat{q}
\end{equation}
The dimension is again $\frac{1}{2}n(n+1)$, and group composition
amounts to adding the $C$ matrices.
\end{list}

Some other subgroups of $Sp(2n,\Re)$ will appear in connection with
global decomposition theorems.

\underline{Global decomposition Theorems}

Now we describe four useful ways of expressing any $S \in Sp(2n,\Re)$
as a product of specially chosen factors, either two or three in
number.

\begin{list}{(\alph{listnumber})}{\usecounter{listnumber}}
\item \underline{Polar Decomposition}~\cite{polar-decomp}: This says that any
$S
\in  Sp(2n,\Re)$ can be written uniquely as the product of two
factors, one belonging to the maximal compact subgroup $K(n)$, the
other to an important sub set $\Pi(n)$ in $Sp(2n,\Re)$.  This subset
is defined by
\begin{equation}
\Pi(n)  =  \{ S \in  Sp(2n,\Re) \mid  S^{T} =  S,  S
\quad\mbox{positive definite} \}  \subset Sp(2n,\Re)
\end{equation}
and it is definitely not a subgroup.  The decomposition reads:
\[
S \in Sp(2n,\Re):\quad S = S(X,Y)P \quad{\rm uniquely},
\]
\begin{equation}S(X,Y)  \in  K(n), \quad P \in  \Pi(n)
\end{equation}
Of course by conjugating $P$ with $S(X,Y)$ one could have written the
two factors in the opposite sequence.  The important points here are
the global nature of this result, and the uniqueness of the factors.
{}From this decomposition one can see that of the two possibilities
$det \; S = \pm 1$ allowed by eqn (2.5) the choice $det \; S = +1$ is
the only one allowed.
\item \underline{Euler Decomposition}:  Next we turn to a
decomposition which involves three factors, each drawn from a
subgroup of $Sp(2n,\Re)$, but which is nonunique.  Two of the factors
are from $K(n)$, the third from those elements of $\Pi(n)$ which are
diagonal and do form a subgroup:
\begin{eqnarray}
S \; \in \; Sp(2n,\Re) \quad S & = & S(X_{1},Y_{1}) \;
D(\b{$\kappa$}) \; S(X_{2}, Y_{2}),
\nonumber \\
S(X_{1}, Y_{1}), \; S(X_{2}, Y_{2}) \; &\in &\; K(n),
\nonumber \\
D(\b{$\kappa$}) & = & diag(\kappa_{1}, \cdots, \kappa_{n},
\kappa_{1}^{-1},
\cdots,\kappa_{n}^{-1}) \; \in \; \Pi(n),
\nonumber \\
\kappa_{r} & > & 0, \quad r \; = \; 1, \cdots, n
\end{eqnarray}
If one adds the numbers of free parameters in the three factors, one
gets the sum $n(2n+1)$ which is just the dimension of $Sp(2n,\Re)$.
Thus the nonuniqueness of this decomposition is of a discrete, not a
continuous, nature. It stems essentially from the freedom to order
the first $n$ diagonal elements of $D(\b{$\kappa$})$ in any way we
like.
\item \underline{Pre-Iwasawa Decomposition}: Here we have a
three-factor unique decomposition which one encounters on the way to
establishing the (next) Iwasawa decomposition but the factors do not
all belong to subgroups of $Sp(2n,\Re)$.  The present decomposition
results from attempting to reduce the off diagonal block $B$ in a
general $S \; \in \; Sp(2n,\Re)$ to zero, by using an element of
$K(n)$ on the right.  The result reads:
\begin{eqnarray}
S \; = \;\left( \begin{array}{cc} A & B \\ C & D \end{array} \right)
\; \; &=&
\; \;
\left( \begin{array}{cc} 1 & 0 \\
C_{o}A_{o}^{-1} & 1 \end{array} \right)\; \;
\left( \begin{array}{cc} A_{o} & 0 \\
0 & A_{o}^{-1} \end{array} \right) \; \;
\left( \begin{array}{cc} X & Y \\
-Y & X \end{array} \right),
\nonumber \\
A_{o} &=& (AA^{T} + BB^{T})^{1/2},
\nonumber \\
X - iY \; &=& \; A_{o}^{-1}(A-iB),
\nonumber \\
C_{o} \; &=& \; (CA^{T} + DB^{T})A_{o}^{-1}
\end{eqnarray}
Here the matrix $A_{o}$ is to be chosen symmetric positive definite,
and all factors are unique.  The symmetry of $C_{o} \; A_{o}^{-1}$
can be checked, so the first factor lies in the lens subgroup $T^{l}$
of eq.(4.17).  The middle factor belongs to the intersection
$GL(n,\Re) \; \cap \; \Pi(n)$, which is not a subgroup.  And the
third factor is from $K(n)$.  This particular decomposition is of
importance in obtaining the generalised Huyghens kernel, which we
describe in Section 6.
\item \underline{Iwasawa Decomposition}~\cite{iwasawa-decomp}:  The polar and
pre-Iwasawa decompositions are similar in that they involve unique
factors, but each factor is not taken from a subgroup.  The Euler
decomposition solves the latter problem, but in the process
uniqueness is lost.  The fourth and last Iwasawa decomposition
retains both virtues: it is global, has unique factors, and each is
taken from a characteristic subgroup of $Sp(2n,\Re)$.  It is a result
of fundamental group theoretical significance, valid for all simple
noncompact Lie groups.  The three subgroups involved are the maximal
compact $K(n)$, a certain maximal Abelian subgroup ${\cal A}$, and a
certain nilpotent subgroup ${\cal N}$.  Therefore this decomposition
is often called the ${\cal KAN}$ decomposition.  We first display it
for $Sp(2,\Re)$:
\begin{eqnarray}
S & = & \left( \begin{array}{cc} a & b \\ c & d \end{array} \right)
\quad \in \quad Sp(2,\Re), \quad ad-bc \; = \; 1:
\nonumber \\
S & = & \left( \begin{array}{cc} 1 & 0 \\ \xi & 1 \end{array}
\right) \; \; \left( \begin{array}{ll} e^{\eta/2} & 0 \\
0 & e^{-\eta/2} \end{array}
\right) \; \; \left( \begin{array}{ll} Cos \varphi/2 & -Sin \varphi/2 \\
Sin \varphi/2 & Cos
\varphi/2 \end{array} \right),
\nonumber \\
\xi & = & (ac + bd) / (a^{2} + b^{2}) \quad \in \quad
(-\infty, \infty),
\nonumber \\
\eta & = & ln(a^2+b^2) \quad \in \quad (-\infty,\infty)\; ,
\nonumber \\
\varphi & = & 2 {\rm arg} (a-ib) \in (-2\pi,2\pi]
\end{eqnarray}
Here the first $- \xi -$ factor belongs to the subgroup ${\cal N}$,
coinciding for $n=1$ with the lens subgroup $T^{l}$; the second $-
\eta -$ factor belongs to the subgroup
${\cal A}$; and the third $- \varphi -$ factor is from $K(1) \; = \;
SO(2)$.

For general $Sp(2n,\Re)$, the situation is more involved.  The
subgroups ${\cal A}\/$ and ${\cal N}\/$ are:
\begin{eqnarray}
{\cal A} & = & \left\{ D(\b{$\kappa$}) \; = \; diag \; (\kappa_{1},
\cdots, \kappa_{n}, \kappa_{1}^{-1}, \cdots,
\kappa_{n}^{-1}) \; \mid \; \kappa_{r}
\; > \; 0 \right\} \; \subset \; \Pi(n);
\nonumber \\
{\cal N} & = & \left\{ \left( \begin{array}{ll} A & \quad 0 \\ C &
(A^{-1})^{T} \end{array} \right) \; \Bigm\vert \; A \; = \;
\left( \begin{array}{ccc} 1 &\cdots & 0 \\
\vdots &1_{\ddots} & \\ &&1 \end{array} \right),
\quad A^{T}C \; {\mbox symmetric} \right\} \subset \; Sp(2n,\Re)
\end{eqnarray}
The abelian subgroup ${\cal A}$ consists of just the elements
$D(\b{$\kappa$})$ that were used in the Euler decomposition (4.20).
The Iwasawa decomposition for $Sp(2n,\Re)$ then states that any $S \;
\in \; Sp(2n,\Re)$ can be uniquely expressed as the product of
three factors,
\begin{equation}
S \; = \; \left( \begin{array}{ll} A & \quad 0 \\ C & (A^{-1})^{T}
\end{array}
\right) \quad D(\b{$\kappa$})
\;\;\;\; S(X,Y)
\end{equation}
taken respectively from ${\cal N, A}\/$ and $K(n)$.  The
dimensionalities of these subgroups, respectively $n^{2}\/$, $n\/$
and $n^{2}$, add up correctly to $n(2n+1)$.
\end{list}

\section{\underline The Lie algebra of $S\lowercase{p}(2\lowercase{n},\Re)$}

We first study the Lie algebra \underline{$Sp(2n,\Re)$} in the
defining representation, and then generalise to any other
representation.  In keeping with quantum mechanical convention, we
shall retain a factor of $i$ in the definition, even though this
might seem unnecessary in dealing with a group of real matrices.

We examine the form of matrices $S \in Sp(2n,\Re)\/$ close to the
identity:

\begin{eqnarray}
S = \exp{(-i\epsilon J)} &\simeq& 1-i\epsilon J, \quad
\vert\epsilon\vert << 1 :
\nonumber \\
S \beta S^{T} = \beta &\Rightarrow& (\beta J)^{T} = \beta J, (J
\beta)^{T} = J \beta,
\nonumber \\
J^{*} &=& -J
\end{eqnarray}
Thus the generator matrix $J$ is pure imaginary, and both $\beta J$
and $J \beta$ are symmetric.  In other words in the defining
representation we get all possible $J$'s by pre or post multiplying
all possible pure imaginary symmetric $2n \times 2n\/$ matrices by
$\beta$.  Taking the former alternative and choosing the simplest
possible basis for symmetric $2n \times 2n\/$ matrices, we obtain the
following basis for \underline{$Sp(2n,\Re)$}:
\begin{eqnarray}
X_{ab}^{(0)} & = & X_{ba}^{(0)}, \quad a,b = 1, \cdots, 2n;
\nonumber\\
(X_{ab}^{(0)})_{cd} & = & i (\delta_{ad} \beta_{cb} +
\delta_{bd} \beta_{ca})
\end{eqnarray}
These matrices can be easily seen to obey the commutation relations
\begin{equation}
[X_{ab}^{(0)}, X_{cd}^{(0)}] = i(\beta_{ac} X_{bd}^{(0)} +
\beta_{bc} X_{ad}^{(0)} + \beta_{ad} X_{cb}^{(0)}  +
\beta_{bd}X_{ca}^{(0)} )
\end{equation}
The structure of \underline{$Sp(2n,\Re)$} is determined by these
relations.  In a general representation of $Sp (2n,\Re)$ we have
generators $X_{ab} = X_{ba}$ obeying
\begin{equation}
[X_{ab}, X_{cd}] = i(\beta_{ac} X_{bd} + \beta_{bc} X_{ad} +
\beta_{ad} X_{cb} + \beta_{bd} X_{ca})
\end{equation}
Finite dimensional representations of $Sp(2n,\Re)$ are necessarily
nonunitary, hence in them the $X_{ab}$ cannot all be hermitian.  This
is because of the noncompactness of $Sp(2n,\Re)$.  On the other hand,
in a unitary representation which is necessarily infinite
dimensional, we have $X_{ab}^{\dag} = X_{ab}$.

To help identify the subsets of generators for various subgroups it
is useful to use split index notation.  We use $r$, $s$, $\cdots = 1,
\cdots, n$ to label the various canonical pairs; and $\alpha,
\beta, \cdots = 1, 2$ to pick out the q and the p in each pair:
\begin{eqnarray}
a, b, \cdots = 1, \cdots, 2n & : & a \rightarrow r
\alpha, b \rightarrow s \beta;
\nonumber \\
\beta_{ab} & = & \beta_{r \alpha, s \beta} = \delta_{rs}
\epsilon_{\alpha \beta},
\nonumber \\
\epsilon & = & \left( \begin{array}{cc}
0 & 1\\ -1 & 0 \end{array} \right)
\end{eqnarray}
Then the various components of $X_{ab} = X_{r \alpha, s \beta}$ are
handled thus:
\begin{eqnarray}
X_{r1, s1} & = & V_{rs} = V_{sr} ;
\nonumber \\
X_{r1, s2} & = & W_{rs}
\nonumber \\
X_{r2, s2} & = & Z_{rs} = Z_{sr}
\end{eqnarray}
There are $\frac{1}{2} n(n+1)\; V$'s, a similar number of $Z$'s, and
$n^{2}\; W$'s; in a unitary representation, each is hermitian.  In
this split form the commutation relations (5.4) read:
\begin{eqnarray}
\left[W_{rs}, W_{uv}\right] & = & i(\delta_{rv} W_{us} -
\delta_{us} W_{rv}),
\nonumber \\
\left[W_{rs}, V_{uv}\right] & = & -i
(\delta_{us} V_{rv} + \delta_{vs} V_{ru}),
\nonumber \\
\left[W_{rs}, Z_{uv}\right] & = & i (\delta_{ru} Z_{sv} +
\delta_{rv} Z_{su}),
\nonumber \\
\left[V_{rs}, Z_{uv}\right] & = & i (\delta_{ru} W_{sv} + \delta_{su}
W_{rv} + \delta_{rv} W_{su} + \delta_{sv} W_{ru}),
\nonumber \\
\left[V, V\right] & = & \left[Z, Z\right] = 0
\end{eqnarray}
Now one can pick out the subsets of generators for various subgroups
of $Sp(2n,\Re)$; we give the results in the form of a Table.
\begin{center}
\begin{tabular}{lr}
\underline{Subgroup} & \underline{Generators}\\
$GL(n,\Re)$ & $W_{rs}$\\ $SO(n,\Re)$ & $J_{rs} = W_{sr} - W_{rs}$\\
$U(n) = K(n)$ & $J_{rs}, Q_{rs} = V_{rs} + Z_{rs}$\\ & $A_{rs} =
\frac{1}{2} (Q_{rs} - i J_{rs})$\\
$T^{(f)}$ & $Z_{rs}$\\ $T^{(l)}$ & $V_{rs}$\\ ${\cal A}$ & $W_{rr}, r
= 1, \cdots, n$\\ ${\cal N}$ & $W_{rs}$ for $r < s$, and all
$V_{rs}$.
\end{tabular}
\end{center}

We have mentioned that any nontrivial finite dimensional
representation of $Sp(2n,\Re)$ is necessarily nonunitary.  It turns
out that, with no loss of generality, we may assume that the
``compact'' generators of $K(n)$ are hermitian, while a balance of
``non compact'' generators are antihermitian.  That is, in any finite
dimensional representation we can assume the following:

\begin{eqnarray}
\mbox{Generators  of} \; K(n) & = &\mbox{compact generators}
\nonumber \\
& = & W_{rs} - W_{sr}, V_{rs} + Z_{rs} = \mbox{hermitian};
\nonumber \\
\mbox{Balance of generators} & = & \mbox{non compact generators}
\nonumber \\
& = & W_{rs} + W_{sr}, V_{rs} - Z_{rs} = {\rm antihermitian}
\end{eqnarray}
The noncompact generators can be arranged into complex combinations
with definite tensor behaviour under $U(n)$.  These combinations are
\begin{eqnarray}
T_{rs} & = & T_{sr} = V_{rs} - Z_{rs} - i (W_{rs} + W_{sr}),
\nonumber \\
\overline{T}_{rs} & = & \overline{T}_{sr}  =
V_{rs} - Z_{rs} + i (W_{rs} + W_{sr}).
\end{eqnarray}
Then the complete set of $Sp(2n,\Re)\/$ commutation relations (5.4 )
appears in a $U(n)\/$ adapted form~\cite{tensor-notation}:
\begin{eqnarray}
[A_{rs}, A_{uv}] &=& \delta_{su} A_{rv} - \delta_{rv} A_{us}\;;
\nonumber \\
\mbox{}[A_{rs}, T_{uv}]  &=&  \delta_{su} T_{rv} + \delta_{sv} T_{ru}\;;
\nonumber \\
\mbox{}[A_{rs}, \overline{T}_{uv}] &=& -\delta_{ru} \overline{T}_{sv} -
\delta_{rv} \overline{T}_{su}\;;
\nonumber \\
\mbox{}[T_{rs}, \overline{T}_{uv}]  &=& -4(\delta_{ru} A_{sv} + \delta_{rv}
A_{su} +
\delta_{su} A_{rv} + \delta_{sv} A_{ru})\;;
\nonumber \\
\mbox{} [T, T]  &=&  [\overline{T}, \overline{T}] = 0
\end{eqnarray}
We see that $T_{rs}$ and $\overline{T}_{rs}$ are second rank
symmetric tensors under $U(n)$, of contravariant and covariant types
respectively.  While in any representation we can assume
$A_{rs}^{\dag} = A_{sr}$, only in unitary representations do we have
$T_{rs}^{\dag} = \overline{T}_{rs}$ as well.

\section{The Metaplectic Unitary Representation and Generalised
Huyghens Kernel}

We saw in Section 2 that for each $S \in Sp(2n,\Re)$, on account of
the Stone-von Neumann Theorem, we can construct a unitary operator
${\cal U}(S)\;$ such that eq. (2.8) holds.  Clearly the phase of
${\cal U}(S)$ is free.  We can ask if this $S$-dependent phase can be
chosen so as to make the composition law (2.9) of the ${\cal U}$'s as
simple as possible.  The answer is that this can be done, and upon
maximum simplification we can achieve
\begin{equation}
S_{1}, S_{2} \in Sp(2n,\Re) \quad {\cal U}(S_{1}) {\cal U}(S_{2}) =
\pm {\cal U}(S_{1}S_{2})
\end{equation}
This sign ambiguity cannot be eliminated.  So we say that we have
here a two-valued unitary representation of $Sp(2n, \Re)$.  A more
correct or useful statement is that the operators involved provide a
faithful unitary representation of the metaplectic group $Mp(2n)$,
which is a two-fold covering of $Sp(2n, \Re)$~\cite{metaplectic}.
Strictly speaking this means that the argument of ${\cal U}(\cdot)$
should be an element of $Mp(2n)$, not $S \in Sp(2n, \Re)$.  However,
having made this point, we shall continue to write ${\cal U}(S)$ as
in eqs. (2.8, 2.9, 6.1).

The generators of this metaplectic representation of $Sp(2n, \Re)$
are all hermitian; in terms of $\hat{q}$'s and $\hat{p}$'s they are
the quadratic expressions~\cite{mp-gen}

\begin{eqnarray}
\hat{W}_{rs} & = & \frac{1}{2} \{ \hat{q}_{r}, \hat{p}_{s} \},
\nonumber \\
\hat{V}_{rs} & = & \hat{q}_{r} \hat{q}_{s}, \quad \hat{Z}_{rs}
= \hat{p}_{r} \hat{p}_{s}
\end{eqnarray}
The characteristic differences between the compact and the noncompact
combinations become clear when expressed in terms of $\hat{a}$'s and
$\hat{a}^{\dag}$'s:
\begin{mathletters}
\begin{eqnarray}
\mbox{Compact generators :}\;
\hat{W}_{rs} -\hat{W}_{sr} & = & i(\hat{a}_{s}^{\dag} \hat{a}_{r} -
\hat{a}_{r}^{\dag} \hat{a}_{s}), \nonumber \\ \hat{V}_{rs} +
\hat{Z}_{rs} & = & (\hat{a}_{r}^{\dag} \hat{a}_{s}
+ \hat{a}_{s}^{\dag} \hat{a}_{r} + \delta_{rs}), \\
\mbox{Non compact generators :}\;
\hat{W}_{rs} +
\hat{W}_{sr} & = & i(\hat{a}_{r}^{\dag} \hat{a}_{s}^{\dag} -
\hat{a}_{r} \hat{a}_{s}), \nonumber \\ \hat{V}_{rs} -
\hat{Z}_{rs} & = & \hat{a}_{r}^{\dag} \hat{a}_{s}^{\dag} +
\hat{a}_{r} \hat{a}_{s}.  \end{eqnarray}
\end{mathletters}
We see that the compact generators of $U(n)$ conserve ''total photon
number'', thus this sub group of $Sp(2n, \Re)$ consists of
``passive'' transformations.  The noncompact generators on the other
hand do not conserve ''photon number'', so we may call them
``active'' generators.  These properties are expressed thus:
\begin{eqnarray}
\hat{N}  &=&  \hat{a}^{\dag}_{r} \hat{a}_{r}:
\nonumber \\
\mbox{}[\hat{W}_{rs}  -  \hat{W}_{sr} \; {\rm or} \; \hat{V}_{rs}+
\hat{Z}_{rs}, \hat{N}] & = & 0,
\nonumber \\
\mbox{}[\hat{W}_{rs}  +  \hat{W}_{sr} \; {\rm or} \; \hat{V}_{rs} -
\hat{Z}_{rs}, \hat{N}] & \neq & 0
\end{eqnarray}
In fact, single exponentials of $i$ times real linear combinations of
the compact generators give us operators of the form ${\cal U}(S(X,
Y))$; while single exponentials of $i$ times real linear combinations
of the noncompact generators give us operators of the form ${\cal
U}(P),\;P \in \Pi(n)$.  For this reason the latter may be called
``squeezing transformations'' ~\cite{n-mode}~\cite{two-mode}; and the
polar decomposition (4.19) may be read as stating that any
metaplectic unitary transformation is uniquely the product of a
compact passive factor and a non compact active squeeze factor. The
definition and production of squeezed states are taken up in more
detail in Section X.

The Schr\"{o}dinger description of the Hilbert space ${\cal H}$ on
which the metaplectic representation acts has been given in eq.
(2.7).  In this description, the eigenvectors $\mid
\b{$q$} >$ of the commuting position operators $\hat{q}_{r}$
appear as a basis:
\begin{eqnarray}
\vert\psi > \in {\cal H}  : \quad \psi(\b{$q$})  & = &  < \b{$q$} \mid \psi>,
\nonumber \\
\hat{q}_{r} \mid \b{$q$} >  & = &  q_{r} \mid \b{$q$} >,
\nonumber \\
< \b{$q$}^{\prime} \mid \b{$q$} > & = & \delta^{(n)}
(\b{$q$}^{\prime} - \b{$q$}),
\nonumber \\
< \b{$q$} \mid \hat{p}_{r} & = & -i \hbar \frac{\partial}{\partial
q_{r}} < \b{$q$}\mid
\end{eqnarray}

It is useful to know that certain of the operators ${\cal U}(S)$ have
very simple actions on these basis vectors.  We list them below:
\begin{mathletters}
\begin{eqnarray}
S(A) &=& \left( \begin{array}{ll} A & \quad0 \\ 0 & (A^{-1})^{T}
\end{array}
\right), \quad A \in GL(n,\Re)\; :
\nonumber \\
{\cal U}(S(A)) \mid \b{$q$} \rangle &=& \mid det A
\mid^{1/2} \mid A \b{$q$} \rangle,
\nonumber \\
\langle \b{$q$} \mid {\cal U}(S(A))   &=&  \mid det A \mid^{-1/2} \langle
A^{-1} \b{$q$} \mid \; ;
\\
D(\b{$\kappa$}) &=& diag (\kappa_{1}, \cdots, \kappa_{n},
\kappa_{1}^{-1}, \cdots, \kappa_{n}^{-1}), \kappa_{r} > 0:
\nonumber \\
{\cal U}(D(\b{$\kappa$})) &=& exp \left( -i \sum_{r=1}^{n} ln
(\kappa_{r}) \hat{W}_{rr}\right) \; ,
\nonumber \\
{\cal U}(D(\b{$\kappa$})) \mid \b{$q$} \rangle &=& \left(
\prod^{n}_{r=1}
\kappa_{r} \right)^{1/2}.
\mid \kappa_{1}q_{1}, \cdots, \kappa_{n}q_{n} \rangle \; ,
\nonumber \\
\langle \b{$q$} \mid {\cal U}(D(\b{$\kappa$})) &=& (\prod^{n}_{r=1}
\kappa_{r})^{-1/2} \langle \kappa_{1}^{-1}
q_{1}, \cdots, \kappa_{n}^{-1} q_{n} \mid\; ;
\\
L(g) &=& \left( \begin{array}{rr} 1 & 0 \\ -g & 1 \end{array}
\right) \in T^{(l)}\;,\; g^{T} = g \; :
\nonumber \\
{\cal U}(L(g)) &=& exp (- \frac{i}{2} g_{rs} \hat{V}_{rs}),
\nonumber \\
{\cal U}(L(g)) \mid \b{$q$} \rangle &=& exp (- \frac{i}{2} q^{T} g q)
\mid \b{$q$} \rangle
\end{eqnarray}
\end{mathletters}
With the help of these results, and the pre-Iwasawa decomposition for
elements of $Sp(2n,\Re)$ described in Section 4, it turns out to be
possible to calculate the generalised Huyghen's kernel in
n-dimensions without too much effort.  This kernel is the
configuration space matrix element $< \b{$q$} \mid {\cal U}(S) \mid
\b{$q$}^{\prime} >\/$ of the metaplectic unitary
operator ${\cal U} (S)$.  We recall that in the case of one degree of
freedom, this kernel has the following form~\cite{mp-gen}:
\begin{eqnarray}
S & = & \left( \begin{array}{ll} a & b\\ c & d \end{array} \right)
\in Sp(2,\Re)
\nonumber \\
\langle q \mid {\cal U}(S) \mid q^{\prime} \rangle &=&
\left \{ \begin{array}{l}
\frac{e^{-i\pi/4}}{\sqrt{h\mid b \mid}} exp \left[
i(dq^{2} - 2qq^{\prime} + aq^{\prime 2}) / 2 \hbar b \right], b \neq
0; \\ exp \bigl( icq^{2}/2a \bigr) \delta \bigl( \frac{q}{a} -
q^{\prime}
\bigr) / \mid a \mid^{1/2}, \quad b = 0
\end{array} \right .
\end{eqnarray}
$\bigl($ These results are, strictly speaking, valid only if $S$ is
sufficiently close to the identity, the point being that ${\cal
U}(\cdot)$ is actually a representation of $Mp(2)$ and so carries as
argument an element of this group.  Therefore the generalised
Huyghens kernel is not expressible totally in terms of $S \in
Sp(2,\Re) \bigr)$.  We may regard the case $b \neq 0$ as generic.
This generalises nicely to any number of dimensions and we find:
\begin{eqnarray}
S & = & \left( \begin{array}{ll} A & B \\ C & D \end{array} \right)
\in Sp(2n,\Re), \quad det B \neq 0\;:
\nonumber \\
\langle \b{$q$} \mid {\cal U}(S) \mid \b{$q$}^{\prime}
\rangle & = & \frac{e^{-i
n\pi/4}}{h^{n/2}} \cdot \frac{1}{\sqrt{\mid det B \mid}} \times
\nonumber \\
&&exp \left[ \frac{i}{2\hbar} \left\{ q^{T} D B ^{-1} q - 2 q^{\prime
T} B^{-1} q + q^{\prime T} B^{-1} A q^{\prime} \right\} \right]
\end{eqnarray}
The nongeneric case when $det B = 0$ has to be handled carefully -
then the kernel collapses to a lower dimensional expression with a
certain number of delta function factors.  However, having given an
indication of the structure involved, we will not go into any further
details.

\section{$S\lowercase{p}(2\lowercase{n},\Re)$ actions on
Wigner and Diagonal Coherent State Representations}

Let $\hat{\Gamma}$ be any quantum mechanical operator, specified in
the Schr\"{o}dinger representation by its configuration space kernel
$\langle \b{$q$} \mid \hat{\Gamma}
\mid \b{$q$}^{\prime} \rangle$.  We
can see that if $\hat{\Gamma}$ is conjugated by ${\cal U}(S)$ for
general $S \in Sp(2n,\Re)$, the change in the kernel involves an
integral transformation in which the generalised Huygens kernel (6.8)
and its complex conjugate both appear. We can ask whether there is
any other way of specifying or describing $\hat{\Gamma}$ such that
this change takes a simpler form, not requiring any integrations at
all.  Indeed there is, and it is given by the use of the techniques
due to Weyl, Wigner and Moyal (WWM)~\cite{wwm}.  We describe this
aspect, and then go on to another practically important way of
describing operators, namely via the diagonal coherent state
representation, and its behaviour under $Sp(2n,\Re)$.

Let us hereafter set $\hbar = 1$.  From the configuration space
kernel $\langle \b{$q$} \mid \hat{\Gamma} \mid \b{$q$}^{\prime}
\rangle$ of
$\hat{\Gamma}$ we obtain its Wigner distribution or WWM
representative by a partial Fourier transformation :

\begin{eqnarray}
\hat{\Gamma} \rightarrow W{(\xi)} & = & (2\pi)^{-n}
\int d^{n} q^{\prime} \langle
\b{$q$} - \frac{1}{2} \b{$q$}^{\prime} \mid \hat{\Gamma} \mid \b{$q$} +
\frac{1}{2} \b{$q$}^{\prime} \rangle exp (i\b{$q$}^{\prime}\cdot \b{p}),
\nonumber \\
\langle \b{$q$} \mid \hat{\Gamma} \mid \b{$q$}^{\prime}
\rangle & = & \int d^{n}p W
\bigl( \frac{1}{2} (\b{$q$} + \b{$q$}^{\prime}), \b{p} \bigr) exp \bigl( -i
\b{p} . (\b{$q$} - \b{$q$}^{\prime}) \bigr)
\end{eqnarray}
Here $W(\xi)$ is a function on the classical phase space
corresponding to the quantum system, with arguments which are $2n$
classical $c$-number $q$'s and $p$'s.  As seen above, one can recover
the operator $\hat{\Gamma}$ from its WWM representative $W(\xi)$
unambiguously.  Then one finds that under conjugation by the
metaplectic operators ${\cal U}(S)$, the changes in $\hat{\Gamma}$
are very simply expressed in terms of
$W(\xi)~\cite{conjugate-action}$:
\begin{equation}
\hat{\Gamma}^{\prime} = {\cal U}(S)^{-1} \hat{\Gamma} {\cal U}(S)
\Longleftrightarrow
W^{\prime}(\xi) = W(S \xi)
\end{equation}
This behaviour of $W(\xi)$ may in fact be regarded as the key or
characteristic property of the WWM method in quantum mechanics; we
may say that this description of operators is covariant under the
full symplectic group $Sp(2n,\Re)$.

Next we turn to the diagonal coherent state description of operators
$\hat{\Gamma}$~\cite{k-sudarshan}.  For $n$ degrees of freedom the
coherent states are defined as usual by
\begin{eqnarray}
\mid \b{$z$} \rangle & = & exp \left\{ \ - \frac{1}{2} \sum_{r=1}^{n}
\mid z_{r} \mid^{2} + \sum_{r=1}^{n} z_{r} \hat{a}_{r}^{\dag}
\right\} \mid 0 \rangle,
\nonumber \\
\hat{a}_{r} \mid 0 \rangle & = & 0,
\nonumber \\
\hat{a}_{r} \mid \b{$z$} \rangle & = & z_{r}\mid \b{$z$} \rangle, \quad
z_{r} \in C, \b{$z$} = (z_{1}, \cdots, z_{n})
\end{eqnarray}
These are normalised states, no two being orthogonal,and can be
written also as the result of phase space displacement operators
acting on the ground state $\mid 0 \rangle$:
\begin{eqnarray}
\mid \b{$z$} \rangle & = & exp \left\{ \sum_{r=1}^{n} \bigl( z_{r}
\hat{a}_{r}^{\dag} - z_{r}^{*} \hat{a}_{r} \bigr) \right\}
\quad \mid 0 \rangle,
\nonumber \\
\langle \b{$z$}^{\prime} \mid \b{$z$} \rangle & = & exp \left\{ -
\frac{1}{2} z^{\prime \dag} z^{\prime} - \frac{1}{2}
z^{\dag} z + z^{\prime \dag}z
\right\}
\end{eqnarray}
The coherent states form an (over) complete set; the resolution of
the identity
\begin{equation}
1 = \int \prod_{r=1}^{n} \frac{d^{2}z_{r}}{\pi} \mid \b{$z$} \rangle
\langle \b{$z$} \mid
\end{equation}
shows that any vector $\mid \psi \rangle$ can certainly be expanded
using them :
\begin{eqnarray}
\mid \psi \rangle & = & \int \prod_{r=1}^{n} \frac{d^{2} z_{r}}{\pi}
\; \psi(\b{$z$}^{*}) \; \mid \b{$z$} \rangle,
\nonumber \\
\psi(\b{$z$}) & = & \langle \b{$z$}^{*} \mid \psi \rangle = exp \bigl( -
\frac{1}{2} z^{\dag} z \bigr)\;(\mbox{entire analytic
function of \b{$z$}})
\end{eqnarray}
Moreover the overcompleteness allows expansion of any operator
$\hat{\Gamma}$ in the form of an integral over projections to these
states :
\begin{equation}
\hat{\Gamma} = \int \prod_{r=1}^{n} \frac{d^{2}z_{r}}{\pi} \;
\phi(\b{$z$})\mid\b{$z$} \rangle \langle \b{$z$} \mid
\end{equation}
For given $\hat{\Gamma}$, this expansion and the weight function
$\phi(\b{$z$})$ are unique, however the latter could in general be a
distribution of quite a singular kind.

Now we look at the behaviour under $Sp(2n,\Re)$.  It turns out that
the states $\mid \b{$z$} \rangle$ have a simple behaviour only under
the ``passive'' maximal compact subgroup $K(n) \subset Sp(2n,\Re)\/$:
\begin{eqnarray}
U = X - iY \in U(n):&&
\nonumber \\
&&{\cal U} \bigl( S(X,Y) \bigr) \mid \b{$z$} \rangle = \mid U \b{$z$}
\rangle
\end{eqnarray}
This can be traced to the fact that the generators of $K(n)$ involve
only terms of the form $\hat{a}^{\dag} \hat{a}$, as we see in eq.(6.3
a).  On the other hand, the effect of ${\cal U}(P)$, for any $P \in
\Pi(n)$, on $\mid \b{$z$} \rangle$ involves an
integration using a suitable kernel, namely the generalised Huyghens
kernel expressed in the coherent state language~\cite{mobious}, which
for the $Sp(2,\Re)\/$ case can be written in a simple form in terms
of complex $SU(1,1)\/$ parameters $\lambda \;\mu\/$:
\begin{eqnarray}
\lefteqn{
{\cal K}(z,z^\prime;S^c) =\langle z^{\prime}\vert{\cal U}(S)\vert z
\rangle} \nonumber \\
&=\zeta(\varphi/2)^\star\;\exp\left\{-\frac{\textstyle
z^2}{\textstyle 2} -
\frac{\textstyle z^\prime{}^\star{}^2}{\textstyle 2} -
\vert z^\prime\vert^2\right\}\times \nonumber\\
&\left\{\begin{array}{l}
\frac{\textstyle 1}{\textstyle \sqrt{2\;\vert{\em Im}(\mu -
\lambda)\vert}} \;
\sqrt{1 - \frac{\textstyle\lambda + \mu^\star}{\textstyle\lambda^\star +\mu}}
\; \sqrt{1 + \frac{\textstyle\mu\vphantom{^\star}}{\textstyle\lambda^\star}}
\quad\times\\
\exp\left[ \frac{\textstyle 1}{\textstyle 2\;\lambda^\star}
\left\{(\lambda^\star +
\mu )\;z^2 + z\;z^\prime{}^\star + (\lambda^\star - \mu^\star)\;
z^\prime{}^\star{}^2\right\}\right]\quad,\\ {\em Im}\;(\mu -
\lambda)\neq 0\;,\; {\em i.e.} \;,\varphi\neq 2n\pi\quad; \\
\\
\sqrt{\frac{\textstyle 2\vert{\em \bf Re}\;(\lambda + \mu)\vert}
{\textstyle 1 + (\lambda + \mu)\;{\em \bf Re}\;(\lambda +
\mu)}}\;\exp\left[
\frac{\textstyle (z\;{\em \bf Re}\;(\lambda + \mu) + z^\prime{}^\star)^2}
{\textstyle 1 + (\lambda + \mu)\;{\em \bf Re}\;(\lambda + \mu)}
\right]\quad,\\
{\em \bf Im}\;(\mu - \lambda) = 0\;,\; {\em i.e.} \;,\varphi =
2n\pi\;,\; n=-1,0,1,\;\mbox{or}\;2 \quad.\end{array}\right.
\nonumber\\
S &=\left( \begin{array}{ll} a&b \\c&d \end{array} \right)
\in Sp(2,\Re),\;\;S^c =\left( \begin{array}{ll} \lambda &
\mu \\ \mu^{\star}& \lambda^{\star}\end{array}
\right),\;\;\begin{array}{l}
\lambda = \frac{1}{2}(a+d+ic-ib)\\ \mu = \frac{1}{2}(a-d+ib+ic)
\end{array}
\end{eqnarray}
(Here each individual square root is defined to have a positive real
part).

In contrast to the WWM result (7.2), we now have covariance under
$K(n)$ alone:
\begin{equation}
\hat{\Gamma}^{\prime} = {\cal U} \bigl( S(X,Y) \bigr)^{-1}
\hat{\Gamma}\;{\cal U}
\bigl( S(X,Y) \bigr) \Leftrightarrow \phi^{\prime} \b{(z)} = \phi(U
\b{$z$})
\end{equation}
Active elements of $Sp(2n,\Re)$ change $\phi$ in a manner involving a
nontrivial integral transformation.

\section{Quantum Noise matrices and Their
$S\lowercase{p}(2\lowercase{n},\Re)$ behaviour}

Let $\hat{\rho}$ be the density operator of any (pure or mixed)
quantum state.  For simplicity alone let us assume that the means of
$\hat{\xi}_{a}$ vanish:
\begin{equation}
\langle \hat{\xi}_{a} \rangle = Tr (\hat{\rho} \hat{\xi}_{a})
= 0
\end{equation}
Non zero values for these means can always be reinstated by a
suitable phase space displacement.  The variance or noise or second
order moment matrix of the state $\hat{\rho}$ is then defined as
follows~\cite{n-mode}:
\begin{eqnarray}
V & = &\bigl( V_{ab} \bigr) = \left( \begin{array}{ll} V_{1} & V_{2}
\\ V_{2}^{T} & V_{3} \end{array} \right),
\nonumber \\
V_{ab} & = & V_{ba} = \frac{1}{2} \langle \left\{ \hat{\xi}_{a},
\hat{\xi}_{b} \right\} \rangle
\nonumber \\
&=& \int d^{2n} \xi \quad \xi_{a} \xi{_b} W(\xi);
\nonumber \\
\bigl( V_{1} \bigr)_{rs} & = & \langle \hat{q}_{r} \hat{q}_{s} \rangle,
\nonumber \\
\bigl( V_{2} \bigr)_{rs} & = & \frac{1}{2} \langle \left\{
\hat{q}_{r}, \hat{p}_{s} \right\} \rangle,
\nonumber \\
\bigl( V_{3} \bigr)_{rs} & = & \langle \hat{p}_{r} \hat{p}_{s}
\rangle
\end{eqnarray}
This is a real symmetric $2n \times 2n$ positive definite matrix
subject to further matrix inequalities which express the uncertainty
principles (see below).

We must note the compact way in which we are able to express $V_{ab}$
as a phase space integral involving the WWM representative $W(\xi)$
of the density operator $\hat{\rho}$.  This is a consequence of the
general rule~\cite{wwm}
\begin{equation}
Tr \bigl( \hat{\rho} \exp{\left\{ i \xi_{o}^{T} \beta
\hat{\xi} \right\}} \bigr) = \int d^{2n} \xi \quad W(\xi) \exp{
\left\{ i \xi_{o}^{T} \beta \xi \right\}}
\end{equation}
valid for any numerical $\xi_{o}$.  Indeed, one can regard this
property of $W(\xi)$ as being as basic as the symplectic
transformation rule (7.2). A particular case of eq.(8.3), when
$\xi_0=0$, shows that $W(\xi)$ is normalised, because $Tr \hat{\rho}
= 1$: its phase space integral is unity.  But we must remember that
though for hermitian $\hat{\rho}$ the function $W(\xi)$ is real, it
may not be nonnegative everywhere.

The change in the noise matrix $V$ when $\hat{\rho}$ is changed by a
symplectic transformation is now easily obtained by exploiting
eq.(7.2) along with the above expression for $V_{ab}$ in terms of
$W(\xi)$.  We have the extremely simple law, for any $S \in
Sp(2n,\Re)$:
\begin{equation}
\hat{\rho}^{\prime} = {\cal U}(S) \hat{\rho}
{\cal U}(S)^{-1} \Rightarrow V^{\prime} = S V S^{T}
\end{equation}

(Note that in comparison to the change $\hat{\Gamma} \rightarrow
\hat{\Gamma}^{\prime}$ in eq.(7.2), $S$ has been replaced by $S^{-1}$ here).
We may say that $V$ undergoes a symmetric symplectic transformation,
which preserves its symmetry and positive definiteness.

The information contained in $V$ can also be given in complex form
using second order moments of $\hat{a}$'s and $\hat{a}^{\dag}$'s:
\begin{eqnarray}
V^{(c)} & = & \bigl( V_{ab}^{(c)} \bigr) = \left(
\begin{array}{ll} {\cal A} & {\cal B} \\ {\cal B}^{*}
&{\cal A}^{*} \end{array} \right),
\nonumber \\
V_{ab}^{(c)} & = & V_{ba}^{(c)*} = \frac{1}{2} \langle \left\{
\hat{\xi}^{(c)}_{a}, \hat{\xi}^{(c)\dag}_{b}
\right\} \rangle; \nonumber \\
{\cal A}_{rs} & = & {\cal A}_{sr}^{*} = \frac{1}{2} \langle \left\{
\hat{a}_{r}, \hat{a}^{\dag}_{s}
\right\} \rangle, \nonumber \\
{\cal B}_{rs} & = & {\cal B}_{sr} = \langle \hat{a}_{r} \hat{a}_{s}
\rangle
\end{eqnarray}
Thus $V^{(c)}$ is a hermitian $2n \times 2n$ positive definite matrix
subject to further uncertainty inequalities given below.

The relations connecting these two forms of the noise matrix are:
\begin{eqnarray}
V^{(c)}& = & \Omega V \Omega^{\dag} \; :
\nonumber \\
{\cal A} & = & \frac{1}{2} \{V_{1} + V_{3} + i \bigl(V_{2}^{T} -
V_{2} \bigr) \},
\nonumber \\
{\cal B} & = & \frac{1}{2} \left\{ V_{1} - V_{3} + i \bigl(V_{2}^{T}
+ V_{2}
\bigr) \right\};
\nonumber \\
V_{1} & = & \frac{1}{2} \left\{ {\cal A} + {\cal A}^{*} + {\cal B} +
{\cal B}^{*} \right\},
\nonumber\\
V_{2} & = & \frac{i}{2} \left\{ {\cal A} - {\cal A}^{*} - {\cal B} +
{\cal B}^{*} \right\},
\nonumber\\
V_{3} & = & \frac{1}{2} \left\{ {\cal A} + {\cal A}^{*} - {\cal B} -
{\cal B}^{*} \right\}
\end{eqnarray}

Here the matrix $\Omega$ is as given in eq.(4.4).  And in place of
eq. (8.4) we have the equivalent transformation law

\begin{equation}
\hat{\rho}^{\prime} = {\cal U}(S) \hat{\rho} {\cal U}(S)^{-1}
\Rightarrow V^{(c) \prime} = S^{(c)} V^{(c)} S^{(c){\dag}}
\end{equation}

\section{Williamson's Theorem and Uncertainty Principles}

For general $S \in Sp(2n,\Re)$ the transformation law (8.4) for the
noise matrix is \underline{not} a similarity transformation, it is a
similarity only if $S = S(X,Y) \in K(n)$.  Normally we expect that
the diagonalisation of $V$ will require a matrix belonging to the
group $SO(2n,\Re)$.  However, the fundamental Theorem of Williamson
comes to our rescue~\cite{williamsons}.  This theorem is a complete
answer to the question: Given a real symmetric $2n
\times 2n$ matrix $V$, what is the maximum simplification we
can achieve in the form of $V^{\prime} = SVS^{T}$ by allowing $S$ to
vary all over $Sp(2n,\Re)\/$?  For general $V$, the normal or
canonical form of $V^{\prime}$ is not a diagonal form; however the
Theorem shows that in case $V$ is positive (or negative) definite to
begin with, then we can certainly choose an $S$ so that $V^{\prime}$
is diagonal.  By suitable further rescaling and ordering of elements
we can then achieve the following:
\begin{eqnarray}
\mbox{$V$ real symmetric positive definite:}&&
\nonumber \\
S V S^{T} & = & diag \bigl(\kappa_{1}, \cdots, \kappa_{n},
\kappa_{1},
\cdots, \kappa_{n}\bigr),
\nonumber \\
&&\kappa_{1} \leq \kappa_{2} \leq \cdots \quad \leq \kappa_{n},
\nonumber \\
\mbox{Suitable}\; S &\in& Sp(2n,\Re).
\end{eqnarray}
We shall call this the \underline{Williamson normal form of $V$}.  In
general, the $\kappa_{r}$ are \underline{not} the eigenvalues of $V$
at all.  Also note that when $V$ is in this form, then we have
$V^{(c)} = V$.

In the Williamson normal form we see that for each canonical pair
$\hat{q}_{r}, \hat{p}_{r}$ we have equal uncertainties: $\Delta q_{r}
= \Delta p_{r} = \kappa_{r}^{1/2}$.  Also
\underline{all} the off-diagonal variances vanish. Therefore
the complete statement of the uncertainty principles for all degrees
of freedom would be
\begin{equation}
\kappa_{r} \geq 1/2, \quad r = 1, \cdots ,n
\end{equation}
A given $2n \times 2n$ real symmetric positive definite matrix $V$ is
quantum mechanically realisable as the noise matrix of some state
$\hat{\rho}$ if and only if in its Williamson normal form (9.1) every
diagonal entry is greater than or equal to one-half.  This is an
$Sp(2n,\Re)$ invariant statement.  
to express these uncertainty principles directly in terms of $V$
without actually passing to its Williamson normal form~\cite{n-mode}.
There are several ways of doing this and for illustration we quote
just one, expressed both in terms of $V$ and $V^{(c)}$:
\begin{eqnarray}
V + \frac{i}{2} \beta &=& \mbox{hermitian positive semidefinite}\;,
\nonumber \\
V^{(c)} + \frac{1}{2} \Sigma_{3} &=& \mbox{hermitian positive
semidefinite}.
\end{eqnarray}
We emphasize that any $V, V^{(c)}$ obeying these conditions are
quantum mechanically realisable - they are necessary and sufficient.

There is a subtle distinction between the matrix $V$ being brought to
diagonal form, and the matrix $V^{(c)}$ being diagonal.  From the
relations (8.6) among the two sets of submatrices we can see easily:
\begin{mathletters}
\begin{eqnarray}
V {\rm~diagonal} & \Longleftrightarrow & ~V_{1}, V_{3}\;
\mbox{diagonal}\;,
V_{2} = 0\;, \nonumber \\ & \Longleftrightarrow & {\cal A,B}
\;\mbox{real diagonal} \nonumber \\
& \not\Longrightarrow & V^{(c)} {\rm diagonal}; \\ V^{(c)}\;
\mbox{diagonal} & \Longleftrightarrow & {\cal A} \;
\mbox{real diagonal}\;,\;{\cal B} = 0 \nonumber \\
& \Longrightarrow & V_{1}, V_{3} \;\mbox{diagonal}, V_{2} = 0
\nonumber\\
& \Longleftrightarrow & V \;\mbox{diagonal}
\end{eqnarray}
\end{mathletters}
This means that the set of states which have diagonal $V^{(c)}$ is a
subset of the set of states having diagonal $V$: the former is a more
restrictive condition, so fewer states obey it.  In the particular
situation when $V$ is in Williamson normal form, it is not only
diagonal but in addition $V_{1} = V_{3}$; thus, in spite of the
general statement (9.4 a), we do find $V^{(c)}$ also diagonal in that
case. In complex form, the relevant consequence of Williamson's
Theorem for $V^{(c)}$ is that with the transformation law (8.7),
$V^{(c)}$ can be ''diagonalised''.

\section{A U(n)-Invariant multimode squeezing criterion}

The noise matrix $V$ for any state of a single mode system is two
dimensional and has the form
\begin{equation}
V = \left( \begin{array}{ll} (\Delta q)^{2} & \Delta(q,p) \\
\Delta(q,p) & (\Delta p)^{2}
\end{array} \right)
\end{equation}
with obvious meanings for the various matrix elements.  The usual
Heisenberg uncertainty relation (with $\hbar = 1$) which
reads~\cite{uncertainity1}

\begin{equation}
\Delta q \;\Delta p \geq \frac{1}{2}
\end{equation}
can, as is well known, be strengthened to the
statement~\cite{uncertainity2}
\begin{equation}
det V \equiv (\Delta q)^{2} (\Delta p)^{2} - \bigl(
\Delta(q,p) \bigr)^{2} \geq \frac{1}{4}
\end{equation}
This is in fact the content of the conditions (9.3) in this case, and
this is $Sp(2,\Re)$ invariant.

The state $\hat{\rho}$ with variance matrix $V$ is usually said to be
a squeezed state if either one of the two diagonal elements of $V$
(but of course not both) is less than one half.  However this
definition possesses no useful or interesting continuous invariance
at all.  A definition of squeezing possessing invariance under the
maximal compact $U(1)$ or $SO(2)$ subgroup of $Sp(2,\Re)$ is this:
the state $\hat{\rho}$ is squeezed if and only if the lesser of the
two eigen values of $V$ is less than one half.  A state squeezed in
the former sense is squeezed also in the latter sense but not
conversely, so the former is more restrictive:
\begin{equation}
\Delta q \; \mbox{or} \; \Delta p < \frac{1}{\sqrt{2}}
\Rightarrow \mbox{lesser eigenvalue of} \; V < \frac{1}{2}
\end{equation}
The $U(1)$ - invariant squeezing criterion has been used in several
studies.  It is of course clear that we cannot ask for any more
invariance in the squeezing criterion, for example it would be
meaningless to think of an $Sp(2,\Re)$ invariant squeezing criterion.

Motivated by the above, we now describe a criterion for squeezing for
states of $n$ mode systems~\cite{n-mode}.  Suppose that for a given
state $\hat{\rho}$ the noise matrix $V$ already has some diagonal
element less than one half.  Then we say that the state is manifestly
squeezed.  However it may happen that every diagonal element $V_{aa}$
of $V$ exceeds or equals one half, yet squeezing is buried and not
manifest.  We have divided elements of $Sp(2n,\Re)$ into passive
$U(n)$ elements and active $\Pi(n)$ elements.  We would like to have
a definition of squeezing invariant under passive $U(n)$
transformations.  Such a definition is the following:
\begin{eqnarray}
\hat{\rho} \;\mbox{is a squeezed state} \Longleftrightarrow
\bigl( S(X,Y) V S(X,Y)^{T} \bigr)_{aa}
< \frac{1}{2}\;,
\nonumber \\
{\rm some} \; X-iY \in U(n) \;,
\nonumber \\
{\rm some} \; a = 1,2, \cdots, 2n.
\end{eqnarray}
Thus, $\hat{\rho}$ is squeezed if $V$ shows it to be manifestly so,
or if this happens after a suitable passive transformation.

We know that in general $V$ cannot be diagonalised by similarity
transformations within $K(n)$ - we expect to have to use matrices
from the much larger group $SO(2n,\Re)$, which may be noncanonical.
Nevertheless it is remarkable that the above $U(n)$ - invariant
squeezing criterion can be expressed in terms of the eigenvalue
spectrum of $V$:
\begin{eqnarray}
\hat{\rho} \;\mbox{is a squeezed state} &\Longleftrightarrow&
\nonumber \\
l(V)& = & \;\mbox{minimum eigenvalue of}\; V <
\frac{1}{2}
\end{eqnarray}
This is precisely equivalent to the definition (10.5)

Obviously the squeezed or nonsqueezed nature of a state $\hat{\rho}$
is unchanged by passive elements of $Sp(2n,\Re)$ lying within $K(n)$,
since then $V$ undergoes a similarity transformation which leaves
$l(V)$ unaltered.  To change $l(V)$,and so the status of a state, in
either direction, we
\underline{must} use active noncompact elements lying in
$\Pi(n)$, if we wish to do so within the framework of $Sp(2n,\Re)$
transformations. This justifies our using the term ''squeezing
transformation'' for the metaplectic unitary operator $\cal U(P)\/$
for elements $P \in \Pi(n)$.

\section{Some interesting families of variance matrices}

Let us denote by ${\cal S}$ the set of all allowed noise matrices
$V$, i.e., all physically realisable ones obeying the uncertainty
principles:

\begin{equation}
{\cal S} = \left\{ V = 2n \times 2n \;\mbox{real symmetric positive
definite}\; \vert V + \frac{i}{2} \beta \;\mbox{positive
semidefinite} \right\}
\end{equation}

This is an $n(2n+1)$ parameter family.  We have seen that a general
$V \in {\cal S}$ is not diagonalisable by elements within $K(n)$. We
ask: can we characterise the subset ${\cal S}_{K}
\subset {\cal S}\/$ consisting of all noise matrices
which \underline{are} diagonalisable using elements of $K(n)\;$?  The
answer is that this can be done rather elegantly, and for this the
complex form $V^{(c)}$ of $V$ is more convenient~\cite{n-mode}:
\begin{eqnarray}
{\cal S}_{K} &=& \left\{ V \in {\cal S} \vert {\cal A B} = {\rm
symmetric}
\right\} \subset {\cal S}:
\nonumber \\
V &\in& {\cal S}_{K} \Rightarrow S(X,Y) V S (X,Y)^{T} = {\rm
diagonal},
\nonumber \\
{\rm some}\; U &=& X-iY \in U(n)
\end{eqnarray}
The subset ${\cal S}_{K}$ is an $n(n+2)$ parameter family.

There are two further subsets of ${\cal S}_{K}$ which are
interesting~\cite{n-mode}.  We call them the ``hermitian'' family
${\cal S}_{H}$, and the ``Gaussian'' family ${\cal S}_{G}$ - they
are of dimensions $n^{2}$ and $n(n+1)$ respectively.  The definition
of ${\cal S}_{H}$ is motivated by the fact that the general
transformation rule (8.7) for $V^{(c)}$ becomes very simple if we
have a $K(n)$ element:
\begin{eqnarray}
U \in U(n): \quad V^{(c)} & \longrightarrow & S^{(c)}(U) V^{(c)}
S^{(c)} (U)^{\dag},
\nonumber \\
{\cal A} & \longrightarrow & U {\cal A} U^{-1},
\nonumber \\
& {\cal B} \longrightarrow & U {\cal B} U^{T}.  \quad \quad (11.2)
\end{eqnarray}
So if ${\cal B} = 0\/$ to begin with, it remains zero; and ${\cal A}$
being hermitian can be diagonalised by some $U$.  This would then
result in both $V^{(c)}$ and $V$ becoming diagonal.  Thus, including
the uncertainty conditions (9.3), the definition of ${\cal S}_{H}$
is :
\begin{equation}
{\cal S}_{H} = \left\{ V \in {\cal S}_{K} \mid {\cal B} = 0\;,\;
{\cal A} - \frac{1}{2} \cdot 1 \;\;
\mbox{positive semidefinite} \right\} \subset {\cal S}_{K} \subset {\cal S}
\end{equation}
One can verify that \underline{no} member of this family ${\cal S}_{H}$
is squeezed.

Next to the family ${\cal S}_{G}$.  Here the definition involves the
``noncompact'' subset $\Pi(n)$ of $Sp(2n,\Re)\/$ :
\begin{equation}
{\cal S}_{G} = \left\{ V \in {\cal S}_{K} \mid 2 V \in \Pi(n)
\subset Sp(2n,\Re)
\right\} \subset {\cal S}_{K} \subset {\cal S}
\end{equation}
It is a fact that such noise matrices are physically realisable,
i.e., they obey the uncertainty conditions, and are diagonalisable
within $K(n)$.  Further, except for $V =
\frac{1}{2} \cdot 1$, \underline{every other $V \in {\cal S}_{G}$ is
squeezed}.  So we see incidentally that this is the only common
element in ${\cal S}_{H}$ and ${\cal S}_{G}$.  All in all, we have:
\begin{eqnarray}
{\cal S}_H,{\cal S}_G &\subset& {\cal S}_K \subset {\cal S}
\nonumber \\
{\cal S}_{H} \cap {\cal S}_{G} &=& \left\{ \frac{1}{2} \cdot 1
\right\}
\end{eqnarray}

\section{Gaussian Pure States, Gaussian Wigner Distributions}

The generators of the metaplectic unitary representation of
$Sp(2n,\Re)$ are quadratics in $\hat{q}$'s and $\hat{p}$'s.  The most
general centred $n$-mode Gaussian wave function involves a quadratic
in the $q$'s in the exponent. It turns out that the former act in
very nice and compact ways on the latter~\cite{gaussian}.  We
describe the main features briefly in this Section.

A general centred and normalised $n$-mode Gaussian pure state can be
parametrised by two real symmetric $n \times n$ matrices $u$ and $v$,
of which the former is positive definite:
\[
\psi_{(u,v)} \bigl(\b{$q$} \bigr) = \pi^{-n/4} \bigl( det u
\bigr)^{1/4} exp \left\{ - \frac{1}{2} q^{T} \bigl( u+iv \bigr)
q \right\},
\]
\begin{equation}
\int d^{n}q \mid \psi_{(u,v)} (\b{$q$})\mid^{2} = 1
\end{equation}

For $(u,v) = (1, 0)$ we get the ground state of the isotropic
oscillator in $n$-dimensions:
\begin{eqnarray}
\psi_{(1,0)} (\b{$q$}) = \pi^{-n/4} exp \bigl( - \frac{1}{2}
q^{T} q \bigr),
\nonumber \\
\hat{a}_{r} \quad \psi_{(1,0)} = 0.
\end{eqnarray}
The calculation of the WWM representative for $\psi_{(u,v)}
(\b{$q$})$, and of the noise matrix, are easy since only Gaussian
integrals and moments are involved.  The results are:
\begin{eqnarray}
W_{(u,v)} (\xi) & = & (2\pi)^{-n} \int d^{n}q^{\prime}
\psi_{(u,v)} \bigl(\b{$q$}
- \frac{1}{2} \b{$q$}^{\prime} \bigr) \psi_{(u,v)} \bigl( \b{$q$} +
\frac{1}{2} \b{$q$}^{\prime} \bigr)^{*}
exp \bigl(i \b{$q$}^{\prime} \cdot \b{p}
\bigr)
\nonumber \\
& = & \pi^{-n} exp \left\{ - \xi^{T} G (u,v) \xi \right\},
\nonumber \\
G(u,v) & = & \left( \begin{array}{cc} u + v u^{-1}v & \quad v u^{-1}
\\ u^{-1} v & \quad u^{-1}
\end{array} \right)
\nonumber \\
& = & \bigl( S (u,v)^{-1} \bigr)^{T} S(u,v)^{-1}
\nonumber \\
S(u,v) & = & \left( \begin{array}{ll} u^{-1/2} & 0 \\ -vu^{-1/2} &
u^{1/2} \end{array}
\right) \in Sp(2n,\Re)\; ;
\nonumber \\
V(u,v) & = & \frac{1}{2} G(u,v)^{-1} = \frac{1}{2} S(u,v) S(u,v)^{T}
\in {\cal S}_{G}
\end{eqnarray}
While the calculations leading to these results are elementary, it is
worth paying attention to the structures involved.  The WWM
representative $W_{(u,v)}(\xi)$ is expected to be a Gaussian, with a
positive definite parameter matrix $G(u,v)$ in the exponent.  What is
interesting is the factorization of this matrix in terms of an
$Sp(2n,\Re)$ - matrix $S(u,v)$; an added feature is that this
$S(u,v)$ is an example of the product of the first two factors in the
general pre-Iwasawa decomposition (4.21) for any $S \in Sp(2n,\Re)$!.
That the noise matrix $V(u,v)$ should be essentially the inverse of
$G(u,v)\/$ is clear from eq.(8.2); it is then the product structure
for $G(u,v)$ that results in $V(u,v)$ being an element of the family
${\cal S}_{G}$ defined in the previous Section.  Incidentally we
also see that $\Pi(n) \subset Sp(2n,\Re)$ defined in eq.(4.18) and
used in defining the family ${\cal S}_{G}$ in eq.  (11.4) can be
described more explicitly using $S(u,v)$:

\begin{eqnarray}
\Pi(n) = \{ S(u,v) S(u,v)^{T} &\vert& S(u,v) \in
Sp(2n,\Re)\,\nonumber \\ &&\mbox{$u$ and $v$ real symmetric $n \times
n$ matrices, $u$ positive definite} \}
\end{eqnarray}
We can argue further along similar lines, concerning the form to be
expected for the action of an operator ${\cal U}(S)$ on any
$\psi_{(u,v)} (\b{$q$})$.  From eq. (7.2) it is clear that
$W_{(u,v)}(\xi)\/$ must get mapped onto another Gaussian; the fact
that it arises from a pure state wave function must also be retained;
and the transformation rule (8.4) plus the explicit form of $V(u,v)$
in (12.3) means that any $\psi_{(u,v)} (\b{$q$})$ can be mapped onto
$\psi_{(1,0)} (\b{$q$})$ by a suitable ${\cal U}(S)$.  All this is
indeed true. We find that
\begin{eqnarray}
\psi_{(u,v)} & = & {\rm (phase ~ factor)} {\cal U}\bigl(S(u,v) \bigr)
\psi_{(1,0)}\; , \nonumber \\
V(u,v) & = & S(u,v) V(1,0) S(u,v)^{T},
\nonumber \\
V(1,0) & = & \frac{1}{2} \cdot 1
\end{eqnarray}
So $Sp(2n,\Re)$ acts \underline{transitively} on the set of Gaussian
pure states.  Since the subgroup of $Sp(2n,\Re)$ leaving the
particular state $\psi_{(1,0)}$ invariant (apart from phases) is just
$K(n)$ -- $\hat{a}_{r}$ annihilates $\psi_{(1,0)}$, and the
generators of $K(n)$ in the metaplectic representation are of the
form $\hat{a}_{r}^{\dag} \hat{a}_{s}$.  More precisely, the situation
may be described as follows.  One can easily establish that the
behavior of $\psi_{(1,0)}\/$ under $K(n)\/$ is given by:
\begin{eqnarray}
U=X-iY &\in& U(n):\nonumber \\ {\cal U}\left( S(X,Y) \right)
\psi_{(1,0)}
&=& \sqrt{Det\; U}\; \psi_{(1,0)}
\end{eqnarray}
The sign ambiguity here explains the appearance of the metaplectic
group $Mp(2n)$. The stability group of $\psi_{(1,0)}$ - the subgroup
of $Sp(2n,\Re)\/$ leaving this vector strictly invariant- is thus the
$n^2-1$ parameter subgroup $SU(n) \in U(n)$. Correspondingly the
orbit of $\psi_{(1,0)}\/$ under $Sp(2n,\Re)$, made up of the vectors
${\cal U}(S)\psi_{(1,0)}\/$ for all $S \in Sp(2n,\Re)$, is an
$(n(n+1)+1)\/$ parameter family.  It consists of the vectors
${\textstyle e}^{i\alpha}\psi_{(u,v)}\/$ for $0\leq\alpha< 2 \pi\/$
and all allowed $(u,v)$. This orbit is essentially the coset space
$Sp(2n,\Re)/SU(n)$; there is a one to one correspondence between
vectors ${\textstyle e}^{i\alpha}\psi_{(u,v)}\/$ and points in this
space.  Suppressing the phase we can next say that the set of density
matrices $\psi_{(u,v)}\psi_{(u,v)}^{\dagger}\/$ , or equally well the
set of representative vectors $\psi_{(u,v)}\;$, is essentially the
coset space $Sp(2n,\Re)/U(n)$.

Turning next to the effect of ${\cal U}(S)$ on $\psi_{(u,v)}$, we see
that apart from a phase factor it has to result in
$\psi_{(u^{\prime}, v^{\prime})}$ for suitable $u^{\prime}$ and
$v^{\prime}$.  The formula for this change is a beautiful one:

\begin{eqnarray}
{\cal U}(S) \psi_{(u,v)} &=& {\rm (phase ~ factor)}
\psi_{(u^{\prime},
v^{\prime})};
\nonumber \\
\wedge &=& (iu - v)^{-1} \rightarrow \wedge^{\prime} = (iu^{\prime} -
v^{\prime})^{-1}
\nonumber \\
&=& (A \wedge + B) (C \wedge + D)^{-1}
\end{eqnarray}

The Gaussian WWM function in (12.3) arose from a pure state.  Suppose
now we consider a general centred normalized Gaussian phase space
distribution with a general parameter matrix G
\begin{eqnarray}
W_{G} (\xi) &=& \pi^{-n} (det G)^{1/2} exp \bigl(-\xi^{T} G \xi
\bigr),
\nonumber \\
G &=& \mbox{real symmetric $2n \times 2n$ positive definite matrix}.
\end{eqnarray}
The question is: when is this a WWM function corresponding to some
pure or mixed quantum state?  Thanks to Williamson's Theorem, the
answer is elementary~\cite{conjugate-action}.  The noise matrix would
clearly be
\begin{equation}
V = \frac{1}{2} G^{-1}
\end{equation}
So we start from the given Gaussian $W_{G}(\xi)$, pass to its $V$,
then go to the Williamson normal form (9.1) of $V\/$ :
\begin{equation}
V \rightarrow {\rm diag} (\kappa_{1}, \cdots,
\kappa_{n}, \kappa_{1}, \cdots,\kappa_{n}),
\end{equation}
and then demand that each $\kappa_{r}$ be greater than or equal to
one half.  This is a complete necessary and sufficient condition for
$W_{G}(\xi)$ to be a bonafide WWM phase space distribution; as we
have seen, however, this condition can be stated directly without
going to the normal form:
\begin{equation}
W_{G} (\xi)\; \mbox{is a WWM distribution} \Longleftrightarrow G^{-1}
+ i \beta = {\rm hermitian ~positive ~semidefinite}
\end{equation}
\section{Concluding Remarks}
In this review we have tried to convey the main features of the
family of real symplectic groups $Sp(2n,\Re)\/$, and have outlined
some problems in optics and quantum mechanics where they are useful.
Our account has been descriptive and suggestive, omitting detailed
properties of various statements made. We believe that any interested
reader wishing to apply symplectic techniques to any concrete problem
would be well equipped for the purpose, and able to supply necessary
details.
\par
Some general remarks - partly to counter apparently common
misconceptions - may be useful at this stage. The group
$Sp(2n,\Re)\/$ comes in when we define \underline{linear} canonical
transformations on given canonical variables. It is the fact that the
commutation relations and hermiticity are maintained that is
responsible for the existence of unitary operators ${\cal U}(S)\/$
implementing these transformations. In particular, ${\cal U}(S)\/$ is
unitary whether $S \in K(n)\/$, when $S\/$ itself is unitary, or $S
\not\in K(n)\/$, for example $S \in \Pi(n)\/$
in which case $S\/$ is hermitian rather than unitary.
Correspondingly, the 'Hamiltonians' generating these unitary
operators ${\cal U}(S)\/$ are hermitian quadratics in the $\hat q$'s
and $\hat p$'s. For the case of the most general unitary evolution
via a general hermitian Hamiltonian, the action on $\hat q$'s and
$\hat p$'s (equivalently on $\hat a$'s and $\hat a^\dagger$'s )
definitely preserves the canonical commutation relations and
hermiticity properties, but the transformed operators may not be
linear combinations of the original ones.
\par
The maximal compact subgroup $U(n)$ of $Sp(2n,\Re)\/$ has naturally
played an important role in our considerations. The $n$-mode
squeezing criterion described in Section 10 has a built-in $U(n)$
invariance. As a result, for a state with a given variance matrix
$V$, squeezing if present may be manifest (one of the diagonal
elements of $V$ is less than $1/2$) or may be hidden (this happens
only after a suitable $U(n)$ transformation).  This makes it clear
that our squeezing criterion is weaker than the usual one stated
directly in terms of the diagonal elements of $V$, since these would
rule out the hidden case. Correspondingly there are more states which
are squeezed by our criterion than by the usual one, which in any
case has much less invariance built in.
\par
The metaplectic group is essential to describe properly the stucture
of the unitary operators implementing linear canonical
transformations on the canonical variables (in this respect the
notation ${\cal U}(S)$ with $S \in Sp(2n,\Re)\/$ is inadequate - the
argument in ${\cal U}(S)$ should be an element of $Mp(2n)$). The
importance of this group is seen in, for example, the calculation of
geometric phases for cyclic evolution of squeezed states, the
interpretation of the Guoy phase, etc~\cite{guoy}. It has also been
shown elsewhere that the metaplectic group is relevant in setting up
operator Mobius transformations for one degree of
freedom~\cite{mobious}.
\par
Work on extending $U(n)$ invariant notions beyond quadrature
squeezing is in progress and will be reported elsewhere. Thus, for
two-mode systems, one can classify squeezing transformations into
$U(2)$ invariant equivalence classes, study bunching and antibunching
with such invariance etc~\cite{two-mode}. These notions can be
generalized to $n$ mode systems as well.

\end{document}